\documentclass[%
preprint,
 amsmath,amssymb,
pra,
]{revtex4-1}
\usepackage{graphicx}
\usepackage{color}
\usepackage{ulem}
\usepackage[nodate]{datetime}
\def\bd{\begin{displaymath}}
\def\be{\begin{equation}}
\def\ed{\end{displaymath}}
\def\ee{\end{equation}}

\begin{document}

\title{\bf Testing Violations of Lorentz Invariance with Cosmic Rays.}
\author{R. Cowsik$^1$, T. Madziwa-Nussinov$^1$, S. Nussinov$^2$, and U. Sarkar$^3$}
\affiliation{\small {1. Physics Department and McDonnell Center for the Space Sciences\\
Washington University, St. Louis, MO 63130\\
2. School of Physics and Astronomy, Tel Aviv University, Ramat Aviv, Tel Aviv 69978, Israel\\
3. Physical Research Laboratory, Ahmedabad 380009, India}}


\begin{abstract}
\noindent Cosmic rays are the highest energy particles available for our study and as such serve as excellent probes of the effects of Lorentz Invariance Violations, which are expected to increase with energy. This general paradigm is investigated in this paper by studying the effects of such violations within the Coleman-Glashow model in which each particle species may have its own maximum attainable velocity, even exceeding that of light \textit{in vacuo}. The particular focus here is that the muon neutrino may have the maximum speed exceeding that of light. We show that such an assumption leads to the elongation of the decay lifetime of the pion that increases with energy over and above the time dilation effects. We provide a transparent analytical derivation of the spectral intensities of muon neutrinos and muons generated in the Earth's atmosphere by cosmic rays. In this derivation we not only account for elongation of the pion lifetime, but also for the loss of energy by the neutrinos by radiation of the electron-positron pairs through the Cohen-Glashow process, during their propagation. We then compare the theoretical spectra with observations of neutrinos and muons from large instruments like IceCube and BUST to set a limit of $\sim10^{-13}$ on the fractional excess speed of neutrinos over that of light. We also show that the ratio of the spectral intensities of downward and upward moving neutrinos at various angles constitute a diagnostic exclusively for the Cohen-Glashow process, which may be searched for in the IceCube data set. We conclude the paper with several comments, including those related to improvements of these tests when definite signals of GZK neutrinos will be observed.
\end{abstract}

\maketitle

\section{Introduction}

\noindent The study of several exciting aspects of high energy astrophysics and indeed of many subtle aspects of basic physics have been given a boost by the commissioning of large detectors of cosmic ray secondaries such as ANITA, IceCube, Auger, BUST, Kolar Gold Fields, Kamiokande and other experiments with collecting powers of $\sim100$ km$^3$ \cite{Krishnaswamy1982, Krishnaswamy1975, Krishnaswamy1971, Gorham2010, Hoover2010, Becker 2011, Aglietta2000, Gray2011, Aharmim2009, Abbasi2011, Novoseltsev2010, Abraham2008, Abbasi2004, MultiKm3, Fechner2009}. These detectors have already detected $\sim10^9$ cosmic-ray muons of median energy $\sim2\times10^4$ GeV and $\sim10^4$ neutrinos that allow the spectra to be determined up to $\sim10^6$ GeV. The physics input regarding high energy nuclear interactions from accelerators, colliders and other sources help in reliably modeling the propagation of cosmic-rays through the atmosphere and qualitatively account for the observed spectral intensities of the muons and the neutrinos. Comparison of these spectral intensities with the model predictions then allow us to probe into primary cosmic ray composition at high energies and search for effects due to new physics, such as small violations of the Lorentz Invariance that may manifest themselves only at the highest energies. This paper is devoted to such an exercise.

Violation of Lorentz Invariance is studied from two distinct perspectives. The first is exemplified by Michaelson-Morley and Hughes-Drever experiments which test the existence of preferred frames of reference and the anisotropy of Machian type long-range interactions of matter in the laboratory with astronomically distant matter. The extraordinary accuracy achieved in such interactions validated relativistic theories of gravity, especially GR \cite{Will1981,Will2006}. The other perspective is exemplified by the theoretical considerations of Coleman and Glashow \cite{Coleman1997, Coleman1999}, who accept the possible existence of a preferred frame, such as the frame in which the dipole anisotropy of the universal microwave background at 2.7K vanishes. In the preferred frame, the laws of physics are assumed to be invariant under translations and rotations. However they investigate the possibility that different particles could have maximum attainable speeds different from that of light  \textit{in vacuo}, and these speeds could, in principle, exceed that of light by a small amount. Coleman and Glashow have developed a perturbative framework to discuss the violations of Lorentz Invariance (LIV) with terms that are renormalizeable and are invariant under the $\textup{SU} (3)\times\textup{SU}(2)\times\textup{U}(1)$ gauge symmetry of the standard model. Going beyond the standard model, Kostelecky and collaborators have carried out extensive analysis of models where Planck scale physics yields signals in the propagation of photons, neutrinos and other particles that have the potential for being observed in present-day or future experiments \cite{Kostelecky2011a, Kostelecky1989, Kostelecky2011b}. These later papers provide a comprehensive overview of the physics and the observational status of these models: Observations of high-energy gamma rays from distant astronomical sources have also been used to set lower bounds on the energy scale at which quantum gravity effects lead to increase in the velocity of light with energy \cite{Biller1999}. The aim of this paper is to discuss the bounds on LIV derived from cosmic-ray observations based on the formalism developed by Coleman and Glashow \cite{Coleman1997,Coleman1999}. In this context, we may refer to the elegant review of earlier work by Bietenholz \cite{Bietenholz2011}.

We begin by recalling briefly the earlier efforts in the field of cosmic rays to search for the effects of superluminal velocities. An excellent review of the efforts to observe tachyons \cite{Bilaniuk1962} in cosmic ray showers is provided by R. W. Clay \cite{Clay1962}. The air-shower group of the Tata Institute of Fundamental Research pioneered these studies by searching for energetic particles that arrive at the air-shower array some 10-50 $\mu$s before the main shower front of electron-positron pairs, muons and  gamma rays initiated by cosmic ray particles of  $\gtrsim10^6$ GeV \cite{Ramana1971, Wolfendale1973}.

Following the lead given by Coleman and Glashow \cite{Coleman1997}, with specific reference to the present paper, the early bounds on LIV using horizontal air showers were obtained by Cowsik and Sreekantan \cite{Cowsik1999}; detailed comments on this paper may be found in the papers of Coleman and Glashow \cite{Coleman1997, Coleman1999} and of Halperin and Kim \cite{Halperin1999}. This later paper maps the violations of Lorentz invariance into violations of the Equivalence Principle. In a subsequent paper, Cowsik et. al. \cite{Cowsik1999b} have investigated the possibility that if similar effects can induce $\nu_\mu\rightarrow \nu_e+\gamma$, then such a rate is far more strictly bounded. Stecker and Glashow \cite{Stecker2001} discuss the bounds on LIV of electrons based on observations of energetic cosmic rays. Similarly Stecker and Scully \cite{Stecker2005, Stecker2009} have put bounds on LIV in the hadronic sector by consideration of the GZK cut-off \cite{Griesen1966, Zatsepin1966}. Direct observations of the neutrinos from supernova 1987A \cite{Hirata1987, Bionta1987} allowed Stodolsky \cite{Stodolsky1988} and Longo \cite{Longo1987} to set bounds on any excess speed of neutrinos over that of light at the $\sim10^{-8}$ level, a significant improvement over early results at accelerators \cite{Adam2011}. In the context of OPERA experiments \cite{Alexandre2012,Adamson2007, Adam2011, Cacciapaglia2011, Haridass2011, Li2011, Gilles2011, Drago2011, Alexandre2011, Bi2011} several ideas  of interest have been put forward and we reference a few for completeness \cite{Ellis2008, Adam2011, Giudice2012, Hagen1987, Grossman2005, Alfaro2005, Cowsik1999, Kostelecky2011a, Hambye1998a, Hambye1998b, Kostelecky1998, Kostelecky1999, Bluhm2000, Pas2005, Hollenberg2009, Coleman1998, Stecker2005, Maccione2009}. 

The particular focus here is to provide an analytical calculation of the spectral intensity of muons and muon neutrinos arising from the decay of pions produced by cosmic rays in the Earth's atmosphere. In carrying out these calculations, we have included the enhancement of pion life-time and the decrease in the average energy transferred to the neutrino in pion decay due to any posited superluminal motion of the muon neutrino. Secondly, we have included the effect of such a neutrino losing energy by emitting electron-positron pairs during its flight even through vacuum, as pointed out recently by Cohen and Glashow \cite{Cohen2011} in the context of OPERA experiments. 

We will not embark here on the ambitious program of getting the best limits for the rich variety of the LIV modifications for the various particles involved. Instead what we will attempt here is more limited and yet clearly illustrates the potential reach of this approach. Here we will focus on the effects of modifying  the energy-momentum relation for $\nu_\mu$ only to $ E_{\nu} = p_{\nu}  (1+\alpha )$, as suggested by Coleman and Glashow for modeling violations of Lorentz Invariance. Our analysis presented here exclude values of $\alpha$ values down to $\sim10^{-13}$. This is achieved by providing a transparent analytical calculation for the propagation of cosmic rays in the Earth's atmosphere that accurately reproduces the known data when no anomaly is assumed, and then comparing the theoretical spectra for various values of $\alpha$ with the observational data. 

\section{Calculation of the spectral intensities of neutrinos and muons generated by cosmic rays in the atmosphere.}

The earliest calculations of the fluxes of neutrinos and muons in the earth's atmosphere were due to Volkova and Zatsepin in 1961 and Zatsepin and Kuzmin in 1962 \cite{Volkova1961, Zatsepin1962}. This was followed by a slightly more detailed calculation by Cowsik and collaborators in 1963 and in 1966 \cite{Cowsik1963, Cowsik1966}. This later paper also describes the experimental aspects of the detection of these energetic neutrinos with detectors located deep underground. Since then, the calculations have progressively improved with the explicit inclusion of the inelastic cross sections for the production of pions and other particles measured with particle beams at accelerators \cite{Volkova1980,Volkova1980a, Honda2007, Gaisser1988, Gaisser12002}. In this section, we derive analytical formulae for the spectral intensities of muons and neutrinos arising from the decay of pions and include the effects of posited superluminal speeds for the muon neutrinos. There are basically two effects: (1) a progressive lengthening of the pion lifetime  \cite{Cowsik2011} and the reduction in energy transferred to the neutrino in the decay process due to LIV effects \cite{Swordy2001}  and (2) the loss of energy suffered by the neutrino during propagation owing to the emission of electron-positron pairs through the Cohen-Glashow process \cite{Cohen2011}. We will describe these two effects below and present an analytical calculation of the cosmic ray fluxes. 

\subsection{Kinematics of pion decay with superluminal neutrinos.}

\noindent In the discussion of the kinematics of pion decay we make the minimal assumption that only the muon neutrino has a maximum attainable speed exceeding that of light \textit{in vacuo}, and work within the framework of the Coleman-Glashow model for LIV \cite{Coleman1997, Coleman1999}. The analysis below follows closely our earlier work in relation to OPERA results \cite{Cowsik2011}. The key assumptions for this analysis are the following: (1) The relation $\partial E/\partial p = v$, the velocity of the particle, (2) energy-momentum conservation holds, and (3) the positivity of energy for free particles, which excludes tachyons. Since the mass of the muon neutrino is in the sub-eV domain and our considerations are limited to neutrinos generated by cosmic rays at high energies, say above 10 GeV, we may safely neglect the neutrino mass and write
\begin{equation}
\label{eq:1}
E_\nu =p_\nu(1+\alpha )
\end{equation}
where $\alpha$ is the superluminal parameter, a very small quantity with $\alpha \ll 1$. Note that $\alpha=2\delta$, where $\delta$ is the LIV parameter defined similarly by Coleman and Glashow \cite{Coleman1997, Coleman1999}.

The superluminal energy-momentum relation in Eq. (\ref{eq:1}) suppresses the pion decay both through its effect on the matrix element of the decay and through kinematic effects, which become progressively more severe with the increasing energy of the pion. We begin with the description of the kinematic effects: The pions and muons follow the standard mass-energy relation
\begin{equation}
\label{eq:2}
E_i =(p_{i}^{2}+m_{i}^{2})^\frac{1}{2}
\end{equation}
It is convenient to express the momentum four-vector of the particles as 
\begin{equation}
\label{eq:3}
\hat{p}_\pi=(E_\pi , p_\pi, 0, 0) , \textup{ }\hat{p}_\mu=(E_\mu , p_{\mu l}, p_{\mu t}, 0)  \textup{  and }   \hat{p}_\nu=(E_\nu , p_{\nu l}, p_{\nu t}, 0) 
\end{equation}
where the subscripts $l$ and $t$ refer to the longitudinal and  transverse components. We explicitly satisfy momentum conservation by choosing 
\begin{equation}
\label{eq:4}
p_{\nu l} = \eta p_\pi, \textup{  }p_{\mu l} = (1-\eta )p_\pi \textup{  and  }  p_{\nu t} = -p_{\mu t} = p_t
\end{equation}
The equation for conservation of energy now reads 
\begin{equation}
\label{eq:5}
( p_\pi^2 + m_\pi^2 )^\frac{1}{2} = \left [ p_\pi^2 (1-\eta)^2 +  p_t^2  + m_\mu^2\right ]^{\frac{1}{2}}+ \left [ p_\pi^2 \eta^2 +  p_t^2 \right ]^\frac{1}{2}(1+\alpha )
\end{equation}
At cosmic ray energies all the momenta are large compared with the masses of the particles and the square roots in Eq. (\ref{eq:5}) may be expanded keeping only the leading terms. This leads to the relation
\begin{equation}
\label{eq:6}
\frac{m_\pi ^2}{2p_\pi } = \frac{m_\mu ^2 + p_t ^2}{2p_\pi (1-\eta )} +\alpha \eta p_\pi  + \frac{p_t ^2 (1+\alpha )}{2p_\pi \eta} 
\end{equation}
\noindent This equation of energy conservation may be thought of as a relationship between $p_t$ and $\eta$. Accordingly, rearranging the terms, we get 
\begin{equation}
\label{eq:7}
p_t ^2= \eta\left \{\frac{(m_\pi ^2 - m_\mu ^2) - \eta\left [ m_\pi ^2 + 2 p_\pi ^2 \alpha (1-\eta )\right ]}{(1+\alpha) - \alpha \eta} \right \}
\end{equation}
\noindent The minimum and maximum value of $\eta$ are obtained by solving Eq. (\ref{eq:7})  for $p_t = 0$:
\begin{equation}
\label{eq:8}
\eta_{min}= 0;  \textup{    } \eta_{max}\approx \frac{m_\pi ^2 - m_\mu ^2}{m_\pi ^2 + 2 p_\pi ^2 \alpha}
\end{equation}
\noindent It is interesting to note that Eq. \ref{eq:8} implies a maximum energy for the neutrino arising from the pions of arbitrarily high energy for a given value of $\alpha$. This maximum energy is given by 
 \begin{equation}
\label{eq:9}
E_{\nu,max}=\frac{m_{\pi}^{2}-m_{\mu}^{2}}{\left (4m_{\pi}^{2}\alpha  \right )^{\frac{1}{2}}}\approx \frac{0.25}{\sqrt{\alpha}}\approx2.5\times 10^{4}\textup{ }GeV\textup{ for }\alpha =10^{-10}
\end{equation}
This limit is noticeable in Fig. \ref{fig:Etan}.

We next consider the modification of the pion decay matrix element due to VLI effects in the Coleman-Glashow model. We begin by writing the pion-decay matrix element in the standard form:
\begin{equation}
\label{eq:10}
\textbf{M}_{\pi\mu } =\frac{g_{w}^{2}}{8m_{w}^{2}}\left \{ \bar{u} (\mu)\gamma _\alpha (1-\gamma ^5) v(\nu _\mu )\right \} f_\pi  \hat{p}_{\pi }^{\alpha }
\end{equation}
\noindent The symbols $\mu$ and $\nu_\mu$ in the brackets next to the wave functions are introduced to indicate the particles they refer to. The essential change with respect to the standard calculation is that we now have 
\begin{equation}
\label{eq:11}
v(\nu _\mu ) \bar{v}(\nu _\mu ) = \not{\hat{p}} - m _\nu + \alpha \vec{\gamma }\cdot \vec{p}
\end{equation}
\noindent After some simplification we have
\begin{equation}
\label{eq:12}
|\textbf{M}_{\pi\mu }|^2 = \left ( f_\pi \frac{g_{w}^{2}}{8m_{w}^{2}} \right )^2 T
\end{equation}
\begin{equation}
\label{eq:13}
T_{\pi \mu } = [2(\hat{p}_\pi \cdot  \hat{p}_\mu )\left \{ \hat{p}_\pi \cdot \hat{p}_\nu-\alpha (\vec{p}_\pi \cdot \vec{p}_\nu) \right \}-m_{\pi }^{2} \left \{ \hat{p}_\nu  \cdot  \hat{p}_\mu - \alpha (\vec{p}_\nu \cdot \vec{p}_\mu)\right \}]
\end{equation}
\noindent The following simplifications are relevant: 
\begin{eqnarray}
\label{eq:14}
\hat{p}_\pi \cdot  \hat{p}_\mu = \frac{1}{2}\left (  m_{\pi }^2 + m_{\mu }^2\right ) - \alpha(\eta ^2 p_{\pi }^2 +p_t^2 ) \nonumber\\
\left \{ (\hat{p}_\pi \cdot  \hat{p}_\nu) - \alpha (\vec{p}_\pi \cdot  \vec{p}_\nu) \right \} = \frac{1}{2}\left (  m_{\pi }^2 - m_{\mu }^2\right ) + \alpha \left \{ p_t^2 -\eta (1-\eta)p_{\pi }^2 \right \}\\
\left \{ (\hat{p}_\nu \cdot  \hat{p}_\mu) - \alpha (\vec{p}_\nu \cdot  \vec{p}_\mu) \right \} = \frac{1}{2}\left (  m_{\pi }^2 - m_{\mu }^2\right ) -  \alpha  \eta p_{\pi }^2  \nonumber\end{eqnarray}
\noindent Accordingly, the decay rate of the pion maybe written as 
\begin{equation}
\label{eq:15}
\Gamma_{\pi \mu }=\int \frac{(2\pi)^4 }{2E_\pi}|M|^2 \delta^4(\hat{p} _\pi - \hat{p} _\mu - \hat{p} _\nu)  \frac{d^3 \vec{p}_\mu }{(2\pi)^32E_\mu}\cdot \frac{d^3 \vec{p}_\nu }{(2\pi)^32E_\nu}
\end{equation}
\noindent The integration $d^3 \vec{p}_\mu$ is accomplished with the choice of 

\noindent $p_{\mu L} = (1-\eta) p_\pi$,   $p_{\mu t} =-p_{\nu t} \textup{ and } p_{\nu L} =- \eta p_{\pi} $, dictated by the $ \delta^3(\vec{p} _\pi - \vec{p} _\mu - \vec{p} _\nu)$ part of the integral. Suppressing the constants, we are now left with the integral

\begin{equation}
\label{eq:16}
\Gamma_{\pi \mu }=\int \frac{T }{E_\mu E_\nu} \delta (E_\pi - E _\mu - E _\nu)  p_\pi \textup{ }d\eta \textup{ }d\varphi \textup{ }p_t \textup{ }dp_t
\end{equation}

\noindent where we have written explicitly $d^3 \vec{p}_\nu =  p_\pi \textup{ }d\eta \textup{ }d\varphi \textup{ }p_t \textup{ }dp_t$

Noting that the Jacobian $\frac{d(E_\pi - E_\mu - E_\nu )}{d p_{t \nu}}= - \frac{d E_\nu}{d p_{t \nu}} = - \frac{(1+ \alpha)}{2 E_\nu}$ and that $d\varphi$ integrates to $2 \pi$, the decay width of the pion is proportional to 
\begin{equation}
\label{eq:17}
\Gamma_{\pi \mu }  = \int_ {{0}}^{{\eta_{max}}} \frac{p_\pi T d\eta}{(1+ \alpha) \left \{ m_\mu ^2 + p_\pi ^2 (1- \eta)^2 + p_t ^2(\eta)\right \}^\frac{1}{2}}
\end{equation}
\begin{figure} [ht]
\includegraphics[width=12cm]{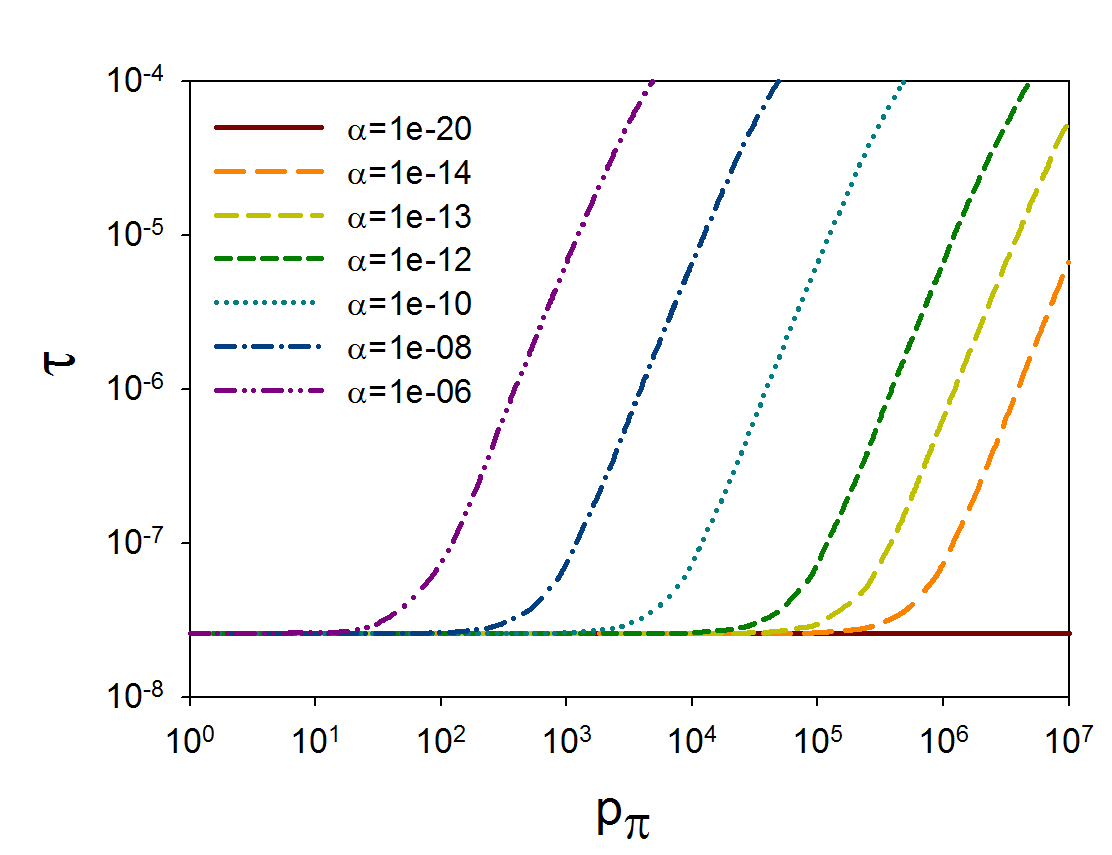}
\caption{We show here the elongation of the decay lifetime of the pion due to superluminal motion of the neutrino, for various values of the parameter $\alpha$ as a function of the pion momentum. All the curves are normalized to $\alpha=0$, for which $\tau$ is taken to be $\sim2.2\times10^{-8}$ s. The additional factor, E$_\pi$/m$_\pi$, is included in the propagation equations, so that for $\alpha=0$ we get the standard results.}
\label{fig:Ptau}
\end{figure}
Note that in Eq. (\ref{eq:17}) $p_t$ is a function of $\eta$ as given in Eq. (\ref{eq:7}), and the limits of the integration are given in Eq. (\ref{eq:8}). As $\eta_{max}$ decreases with increasing $p_\pi$, the decay probability decreases. The other LIV effects are contained in the trace T and the denominator of the integrand in Eq. (\ref{eq:17}). The electronic mode of the pion decay is assumed to be without any LIV effects, and contributes about $10^{-4}$ of the muonic mode for $\alpha=0$. $\Gamma_e \approx 1.2 \times 10^{-4}\textup{ }\Gamma_{\pi -\mu }(\alpha = 0)$ and the full decay width of the pion may be written as 
\begin{equation}
\label{eq:18}
\Gamma_{\pi }= \Gamma_{\pi \mu } + \Gamma_{\pi e }
\end{equation}
\noindent  Noting that the pion lifetime $\tau$ is inversely proportional to $\Gamma$, we show in Fig. \ref{fig:Ptau}. the factor by which the pion life-time is prolonged when $\alpha$ is different from zero at various pion energies.
\subsection{Simple model for Earth's atmosphere.}
\begin{figure}[here]
\includegraphics[width=8cm]{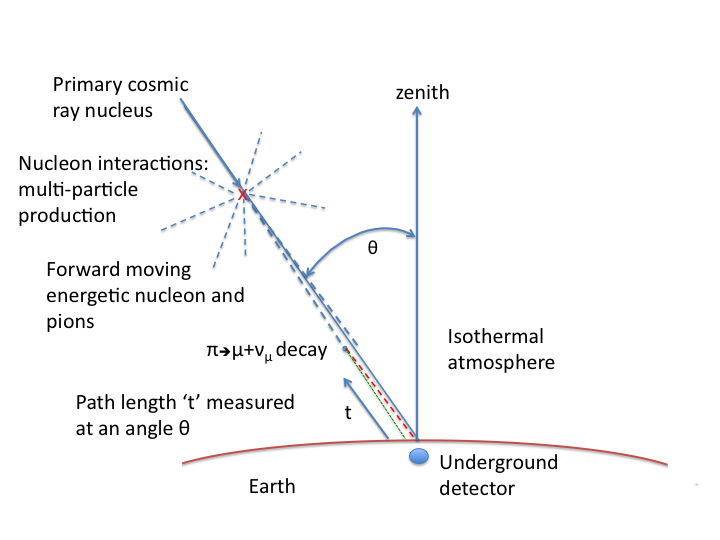}
\includegraphics[width=8cm]{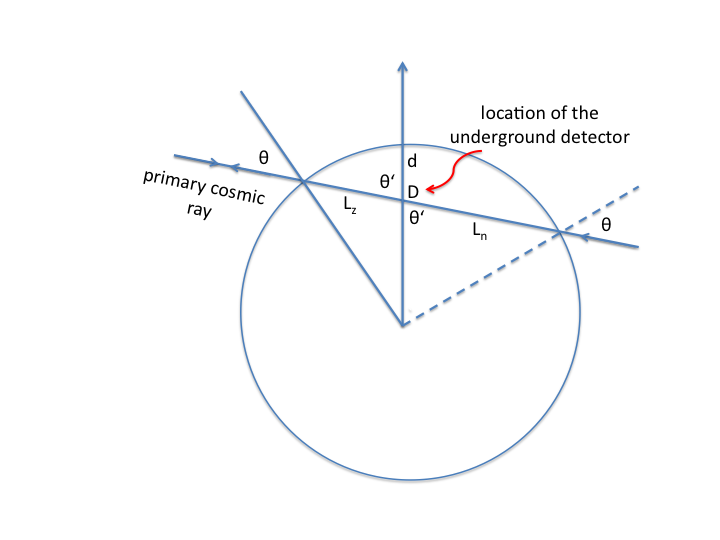}
\caption{The left panel indicates a cosmic ray nucleon incident at a zenith angle $\theta$ that suffers an inelastic interaction with a nucleus in the atmosphere leading to multi-particle production. The surviving energetic nucleon in the atmosphere and the high energy pion travel essentially in the same direction, as do the muons and neutrinos arising from the decay of pions. The path length `$t$' is measured from the earth's surface at the zenith angle $\theta$. The right panel displays the overall geometry: the neutrinos and muons enter the detector D at angle $\theta'\approx \theta$ for the depth d $\ll $ R and $\theta \gtrsim \pi/2$. The zenith angle $\theta$ for downward and upward moving particles through the detectors are the same. }
\label{fig:GeomIO}
\end{figure}
\noindent We  assume an isothermal atmosphere with the density falling off exponentially with height, $h$:
\begin{equation}
\label{eq:19}
\rho(h)= \frac{\textup{ x}_0}{h_0} \textup{ }e^{-\frac{h}{h_0}}
\end{equation}

\noindent Here we take $\textup{x}_0=1030 \textup { g} \textup{ cm} ^{-2}$ and the scale height $h_0=7 \times 10^5 \textup{ cm}$ to express the density $\rho$ in $\textup{g }\textup{cm}^{-3}$. Let $t$ be the length  from the surface of the earth along the path of a cosmic ray incident at a zenith angle $\theta$, as shown in Fig. \ref{fig:GeomIO}. The column density $ \textup{x}(t, \theta)$ that this cosmic ray particle arriving from infinity has to penetrate to reach this point is given by 
\begin{equation}
\label{eq:20}
\textup{x}(t, \theta) =  \textup{x}_0 \textup{ sec} \theta \textup{ }e^{-t/(h_0 sec \theta)} 
\end{equation}
 so that
\begin{equation}
\label{eq:21}
\left | dt/d\textup{x} \right |=h_0 \textup{ } sec \theta/ \textup{x} \equiv H/\textup{x}
\end{equation}
Such an assumption of plane-parallel atmosphere is an adequate approximation up to zenith angles of $\approx  85^o$. (For a cosmic ray particle arriving precisely horizontally with $\theta = \pi /2$, the maximum column density saturates at $\approx 35 \textup{ x}_o$).

\subsection{Propagation of cosmic rays in the Earth's atmosphere.}

\noindent 
At the high energies relevant to the present context, the cosmic rays are incident isotropically on the top of the Earth's atmosphere. They interact repeatedly as they descend into the atmosphere losing energy through the production of secondary particles, mostly pions. These pions are produced with low transverse momenta, $\sim0.5$ GeV/c and consequently the energetic pions and the leading nucleon propagate essentially in the same direction of the primary nucleons. The same is true of the neutrino arising from pion decay where the transverse momentum imparted to the decay products has a maximum of $\approx 45$ MeV/c. Accordingly, we assume all the products of the interactions or decay, including muons generated through the interactions of neutrinos underground preserve the zenith angle of the parent particle. To proceed, we assume a simple power law for the high-energy spectrum of cosmic ray nuclei incident on top of the atmosphere \cite{Swordy2001}. 
\begin{equation}
\label{eq:22}
f_n (E,  \textup{ x}=0,  \textup{ } \theta) = \frac{A_n}{E^{\gamma + 1}} (\textup{ }cm^2 \cdot s \cdot sr \cdot GeV)^{-1}
 \end{equation}
 Here $A_n$ is a constant, $\gamma \approx 1.7$, and $E$ is the energy of the cosmic ray particle in GeV per nucleon. The nucleons interact inelastically with the air nuclei with an effective mean free path, $\lambda_n$ of $\sim80  \textup{ g} \textup{ cm}^{-2}$ generating pions and other particles. The leading nucleon emerges from such collisions with a significant fraction, $\eta _n$, of the initial energy. Because of this, the reduction in flux of primary nuclei in the earth's atmosphere occurs with a mean free path $\Lambda $ that is significantly larger than $\lambda_n$. Specifically it can be shown that
\begin{equation}
\label{eq:23}
 \Lambda = \frac{\lambda_n}{1-<\eta_n^\gamma >}\approx 120\textup{g}/\textup{cm}^2
 \end{equation}
 where
\begin{equation}
\label{eq:24}
<\eta_n^\gamma >=\int_{0}^{1}(\eta_n')^\gamma P(\eta_n') \textup{d}\eta_n'
\end{equation}
with $P(\eta_n') $ the probability that  the leading nucleon emerges with a fraction $\eta'$ of the initial energy. 
We will encounter similar averages; but we will just use the appropriate averages without explicitly showing the angular brackets indicating the average. The flux of the nucleons at a depth x in the atmosphere is then; 
 \begin{equation}
\label{eq:25}
 f_n(E,\textup{x},\theta ) = \frac{A_n}{E^{\gamma +1}}e^{-\textup{x}/\Lambda  }
\end{equation}
These nucleons interact inelastically with air nuclei and generate pions which carry an effective fraction $\eta_\pi$ of the primary energy and have an effective multiplicity $n_\pi$. Thus the rate of production of pions of energy E in the atmosphere at column density depth x is given by
\begin{equation}
\label{eq:26}
q_\pi(E,\textup{x},\theta )=\frac{A_n n_\pi \eta_\pi^\gamma }{\lambda _n E^{\gamma +1}}\cdot e^{-\textup{x}/\Lambda } 
=\frac{B_\pi }{ E^{\gamma +1}}\cdot e^{-\textup{x}/\Lambda }
\end{equation}
The pions  interact and decay in the atmosphere and their spectral intensity is controlled by the equation 
\begin{equation}
\label{eq:27}
\frac{\textup{d}f_\pi(E,\textup{x})}{\textup{dx}} = q_\pi(E,\textup{x},\theta ) - f_\pi(E,\textup{x}) \left \{ \frac{1}{\lambda _\pi} + \frac{m_\pi |\textup{d}t(\theta )/d\textup{x}|}{Ec\tau (\alpha ,E)}\right \}
\end{equation}
Here $\lambda _\pi \sim120$g/cm$^2$ is the interaction mean free path of the pions, and $\tau (\alpha ,E)$ is the pion lifetime elongated due to the LIV $(\alpha\neq 0)$ effects at energy $E$ (see Fig. \ref{fig:Ptau}) and the factor 
$|\textup{d}t(\theta )/d\textup{x}| = h_0 \textup{sec}\textup{ }\theta/\textup{x} $ is the conversion factor from grammage x to path length at zenith angle $\theta$. To facilitate a parallel calculation for the $\alpha = 0$ case, the factor $m_\pi/E$ is shown separately in the decay probability. Defining
\begin{equation}
\label{eq:28}
\varepsilon_\pi=h_0 \textup{ }  \textup{sec } \theta \textup{ } m_\pi /c \tau(\alpha ,E)=Hm_\pi/c\tau(\alpha,E)
\end{equation}
Eq. (\ref{eq:27}) may be written  as
\begin{equation}
\label{eq:29}
\frac{\textup{d}f_\pi(E,\textup{x})}{\textup{dx}} = q_\pi(E,\textup{x},\theta ) - f_\pi(E,\textup{x}) \left \{ \frac{1}{\lambda _\pi } +\frac{\varepsilon _\pi(\alpha ,\theta,E )}{E\textup{x}} \right \}
\end{equation}
For $\textup{sec }\theta=1$ and $\alpha =0$, $\varepsilon_\pi \approx\textup{ }$125 GeV.
The solution to Eq.(~\ref{eq:27}) simplifies considerably for $\Lambda \approx \lambda _\pi \approx $120 g/cm$^2$ to yield
\begin{equation}
\label{eq:30}
f_\pi(E,\textup{x} )= A_\pi\textup{ }\textup{x}\textup{ }e^{-\textup{x}/\lambda _\pi }E^{-(\gamma +1)}\left ( \frac{E}{\varepsilon _\pi +E} \right )
\end{equation}
where $ A_\pi$ is a constant.
\begin{figure}[here]
\includegraphics[width=12cm]{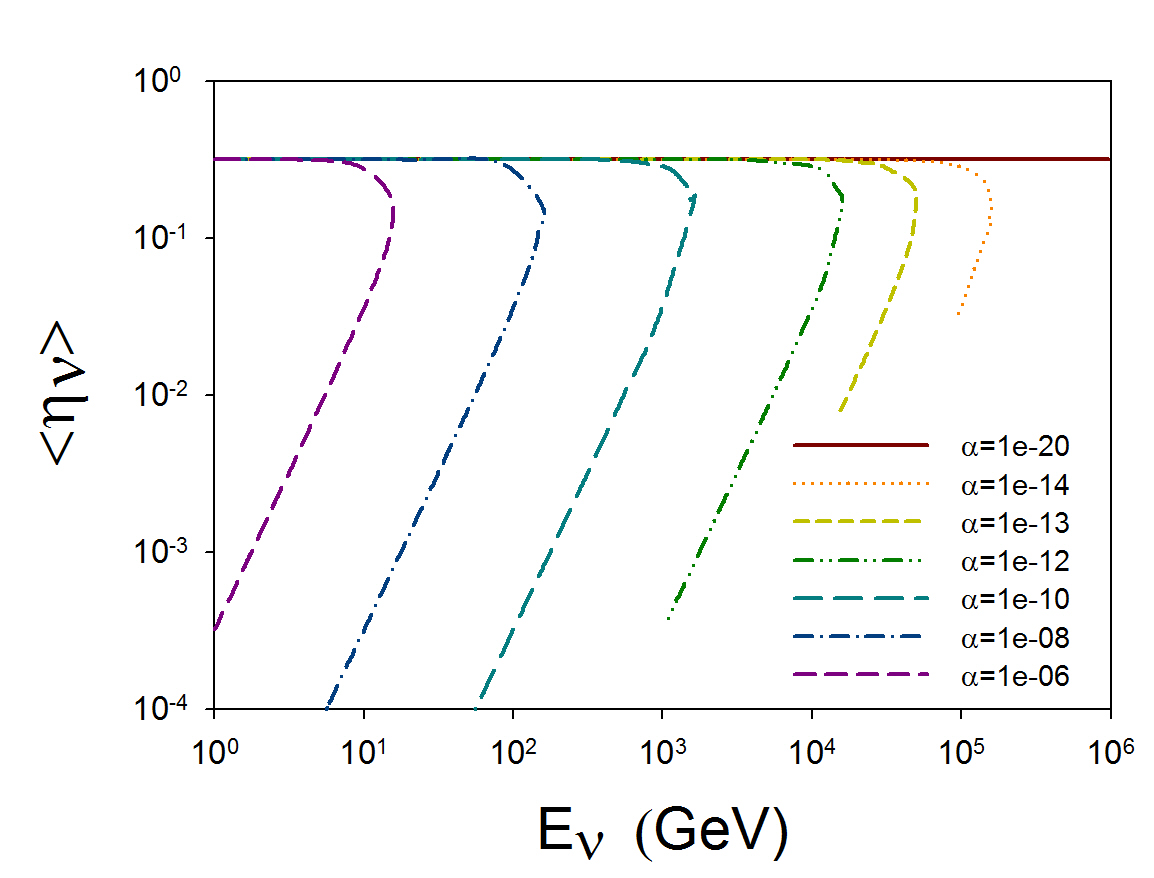}
\caption{The value of $<\eta_\nu>  \textup{ }\approx \textup{ }0.75\textup{ }\eta_{max}$ is shown as a function of $E_\nu$ for various values of $\alpha$. A neutrino of energy E$_\nu$ arises in the decay of a pion of energy E$_\pi=$ E$_\nu/\eta_\nu$. Note that for $\alpha \neq 0$, there are two values of $<\eta_\nu>$ for each $E_\nu$. As the cosmic-ray spectrum is steep, only the larger value of $\eta_\nu$ contributes significantly. Note that for a given value of $\alpha$, there is a maximum value for the neutrino energy, E$_\nu$, as stated in Eq. \ref{eq:9}.} 
\label{fig:Etan}
\end{figure}
The neutrinos arise through the decay of pions in the Earth's atmosphere and the calculation of this production rate involves some subtle considerations. Neutrinos of energy $E$ are produced in the decay of pions with higher energy $E_\pi=E/\eta$. Yet the effective average value of $\eta$ is  in itself a function of $E_\pi$ and $\alpha$. In order to address this issue, we show the weighted average, $<\eta _\nu>$ = $\eta_\nu\approx 0.75 \textup{ }\eta_{max}$ as a function of the energy of the neutrino, $E_\nu$  in Fig. \ref{fig:Etan}. For any given values of the neutrino energy, $E$ and LIV parameter $\alpha$, we then can read off the effective mean value of $\eta$, and find the typical  energy of the pion $E_\pi=E/\eta_\nu$ that generated the neutrino. It is at this energy that we should evaluate the pion lifetime, $\tau (\alpha ,E_\pi =E_{\nu }/\eta _\nu )$, which we write as $\tau_\eta$. Note that for $\alpha \neq 0$ there are two values of $<\eta>$ that occur; because of the steepness of the cosmic-ray spectrum, it is only the larger value of $<\eta>$ that contributes significantly to spectral intensity of the neutrinos. Defining $\varepsilon_\eta =h_0\textup{ sec }\theta \textup{ }m_\pi /(c\textup{ }\tau _\eta)$,  and using a similar reasoning to that used above in deriving equation (\ref{eq:29})and (\ref{eq:30}), the rate of neutrino  production may be written as
\begin{equation}
\label{eq:31}
q_\nu(E,\textup{x})=A_\pi \textup{x }e^{-\textup{x}/\lambda _\pi}E^{-(\gamma +1)}\eta _{\nu }^{\gamma }\left \{ \frac{(E/\eta_\nu)}{\varepsilon _{\eta_\nu} +(E/\eta _\nu)} \right \}\textup{ }\frac{1}{\textup{x}}\textup{ }\frac{\varepsilon _{\eta}}{(E/\eta _\nu)}
\end{equation}
or
\begin{eqnarray}
\label{eq:32}
q_\nu(E,\textup{x})&=&A_\pi \textup{ }e^{-\textup{x}/\lambda _\pi}E^{-(\gamma +1)}\eta _{\nu }^{\gamma +1}\left \{ \frac{\varepsilon _\eta }{\eta _\nu \varepsilon _\eta +E} \right \} \nonumber\\
&\equiv& Q_\nu(E) \textup{ }e^{-\textup{x}/\lambda _\pi}
\end{eqnarray}
where we factored the energy dependent and x dependent terms as $Q_\nu(E)$ and $e^{-\textup{x}/\lambda _\pi}$.

The final step in the calculation of the neutrino flux due to the source function $q_\nu$ is to include the Cohen-Glashow process of energy loss for the neutrinos \cite{Cohen2011}. Noting that in a single emission of an electron-positron pair through this process, the neutrino loses more than $70\%$ of its energy, we treat this process as a decay with an effective lifetime $\tau_G$ given by 
\begin{eqnarray}
\label{eq:33}
\tau_G=\left | E/c (dE/d\textup{x})_{CG} \right | =E/\left \{ c\omega G_{F}^{2} \alpha^{3}E^{6}\right \} \equiv \frac{\tau _{CG}}{E_{GeV}^{5}\alpha ^3} 
\end{eqnarray}
Using the constant $\omega$ given by Cohen and Glashow \cite{Cohen2011}, we find $\tau _{CG} \approx 6.5 \times 10^{-11}s$. Keeping in mind that in writing the differential equation for the evolution of $f_\nu(E,\textup{x})$ we need to introduce the factor $|dt/d\textup{x}|$ for converting grammage to length, as we did before in  Eq.(~\ref{eq:27}) while describing the decay of pions, we write 
\begin{eqnarray}
\label{eq:34}
\frac{\textup{d}f_\nu}{\textup{dx}} &= &q_\nu(E,\textup{x} ) - f_\nu\textup{ }\cdot \frac{h_0\textup{ sec }\theta }{\textup{x}c\tau _G} \nonumber\\
                                                          & =& Q_\nu(E)e^{-(\textup{x}/\lambda _\pi)}-f_\nu \cdot \frac{H}{\textup{x}c\tau _G}
\end{eqnarray}
\noindent To solve this we let 
\begin{eqnarray}
\label{eq:35}
g(E,\textup{x})&=&f_\nu(E,\textup{x})\textup{exp}\left [- \int_{\textup{x}}^{\textup{x}_{max}} \frac{H}{\textup{x}'c\tau _G}d\textup{x}'\right ] \nonumber\\
&=&f_\nu(E,\textup{x}) \left ( \frac{\textup{x}}{\textup{x}_{max}} \right )^{\frac{H}{c\tau _G}}
\end{eqnarray}
where $\textup{x}_{max}=\textup{x}_0\textup{sec}\theta $ and get
\begin{equation}
\label{eq:36}
\frac{\textup{d}g}{\textup{dx}} = Q_\nu(E)\left ( \frac{\textup{x}}{\textup{x}_{max}} \right )^{\frac{H}{c\tau _G}}e^{-\textup{x}/\lambda _\pi}
\end{equation}
This leads to 
\begin{eqnarray}
\label{eq:37}
g &=& Q_\nu(E)\left ( \frac{\textup{1}}{\textup{x}_{max}} \right )^{\frac{H}{c\tau _G}}\int_{0}^{\textup{x}_{max}}\textup{x}^{\frac{H}{c\tau _G}}e^{-\textup{x}/\lambda _\pi}d\textup{x} \nonumber\\
&=&Q_\nu(E)\lambda _\pi \left ( \frac{\lambda _\pi}{\textup{x}_{max}} \right )^{\frac{H}{c\tau _G}}\int_{0}^{\textup{x}_{max}/\lambda _\pi}u^{\frac{H}{c\tau _G}}e^{-u}du \\
&\approx& Q_\nu(E)\lambda _\pi \left ( \frac{\lambda _\pi}{\textup{x}_{max}} \right )^{\frac{H}{c\tau _G}}\Gamma \left (  {\frac{H}{c\tau _G}}+1\right ) \nonumber
\end{eqnarray}
The final step of writing the integral as a complete gamma function follows by noting that $\textup{x}_{max}\gg \lambda _\pi$.
\begin{eqnarray}
\label{eq:38}
f_\nu(E,\textup{x})&=&g(E,\textup{x})\cdot\left ( \frac{\textup{x}_{max}}{\textup{x}} \right )^{-\frac{H}{c\tau _G}} \nonumber\\
&=&A_\pi \lambda_\pi \eta _{\nu}^{\gamma +1}   \left ( \frac{\lambda_\pi }{\textup{x}} \right )^{\frac{h_0\textup{ sec}\theta }{c \tau_G}}  E^{-(\gamma +1)}  \left ( \frac{\varepsilon _\eta }{\eta _\nu \varepsilon _\eta +E} \right ) \Gamma \left ( \frac{H}{c \tau_G}+1 \right )
\end{eqnarray}
At the surface of the earth, $\textup{x}= \textup{x}_{max}$ and the spectral intensity of the neutrinos is given by 
\begin{equation}
\label{eq:39}
f_\nu(E,\textup{x}_{max})=A_\pi \lambda_\pi \eta _{\nu}^{\gamma +1}   \left ( \frac{\lambda_\pi }{\textup{x}_0\textup{sec}\theta } \right )^{\gamma +1}  E^{-(\gamma +1)}  \left ( \frac{\varepsilon _\eta }{\eta _\nu \varepsilon _\eta +E} \right ) \Gamma \left ( \frac{h_0\textup{sec}\theta }{c \tau_G}+1 \right )
\end{equation}
Detectors for cosmic rays and cosmic neutrinos are placed underground to reduce the background due to other particles and gamma rays generated by cosmic rays. Consider such a detector, $D$, placed at a vertical depth, $d$, as shown in the Fig. \ref{fig:GeomIO}. The straight line though $D$ at a zenith angle $\theta' \approx \theta$ (for $\theta \lesssim \pi/2$) emerges from the Earth's surface at distances $L_z$ and $L_n$ respectively. A theorem in Euclidian geometry yields
\begin{equation}
\label{eq:40}
L_z \cdot L_n= (2R-d)d 
\end{equation}
or 
\begin{equation}
\label{eq:40b}
L_n = (2Rd-d^2)/L_z \nonumber
\end{equation}
and $L_z$, for $\theta \lesssim \pi/2$ is given by 
\begin{equation}
\label{eq:41}
L_z =\left |(R-d)\textup{cos}\theta -\sqrt{(R-d)^2\textup{cos}^2\theta +2Rd-d^2} \right |
\end{equation}
From this we can calculate $L_n$ using  Eq. (\ref{eq:40}). The propagation of the neutrino spectral intensities is straightforward if we assume their flux is not significantly depleted due to interactions and assume only the Cohen-Glashow process to operate. Accordingly, their flux at depth $d$ maybe written as 
\begin{eqnarray}
\label{eq:42}
f_\nu(E,\theta,L_z)&=&f(E,\textup{x}_{max},\theta)\textup{exp}\left[-L_z/c\tau_G\right]  \nonumber\\
f_\nu(E,\theta,L_n) &=& f(E,\textup{x}_{max},\theta )\textup{exp}\left[-L_n/c\tau_G\right ] 
\end{eqnarray}
We note that the ratio of these two spectral intensities $R_{z,n}(E)$ is given by
\begin{equation}
\label{eq:43}
R_{z,n}(E)=\textup{exp} \left [ -(L_z-L_n)/c\tau_G(\alpha,E) \right ]
\end{equation}
\noindent It is interesting to add a comment here that for most neutrino telescopes operating underground, the geometrical collecting factor isessentially independent of the hemisphere from which the particle arrives, i.e it is the same for downward and upward moving particles. To the extent we can neglect neutrino oscillation effects, $R_{z,n}(E)$ will be a good probe of the the Cohen-Glashow process. Even though the zenith angle $\theta$, in the Earth's atmosphere, of the particles entering the detector after traversing the distances $L_z$ and $L_n$ are the same, there could be a few percent differences in the scale height of the atmosphere at the antipodal points. When the observational data are averaged over a year, the differences will be reduced further. More importantly, the cross section for the interaction of neutrinos of energy greater than $\sim10^5$ GeV is $\sim10^{-34}(E/10^5 \textup{ GeV})^{\frac{1}{2}} \textup{}\textup{cm}^2$, so that the interaction probability across the diameter of the Earth is about $\sim 30\%$ and increases with increasing neutrino energy as $E^{\frac{1}{2}}$. Allowance for this has to be made during the analysis of the data while searching for the Cohen-Glashow effect. Alternatively, with stringent bounds on $\alpha$ obtained from other observations, the asymmetry in the downward and upward intensities may be used to estimate neutrino cross-sections at high energies. The vacuum oscillation length in meters is $\sim2.5\textup{ }E_\nu\textup{ (GeV)}/\Delta m^2\left ( eV^2 \right )$ so that at the energies of neutrinos under consideration this process may be neglected. 

\subsection{Calculation of the spectral intensities of muons.}

\noindent This calculation follows along similar lines as that for the neutrinos and becomes indeed simpler when we neglect effects of energy loss due to bremsstralung and ionization in the Earth's atmosphere in the region of interest. The muons are generated at a rate
\begin{equation}
\label{eq:44}
q_\mu(E,\textup{x},\theta )=A_\pi\textup{x}e^{-\textup{x}/\lambda _\pi}E^{-(\gamma +1)}\eta _{\mu}^{\gamma } \left \{ \frac{E/\eta _\mu}{\varepsilon _{\eta_\mu}+E/\eta _\mu} \right \} 
\frac{1}{\textup{x}} \frac{\varepsilon _{\eta_\mu}}{E/\eta _\mu}
\end{equation}
Here $\eta_\mu$ is the effective average of fraction of energy that the muon receives in the decay of a pion of energy $E/\eta_\mu$. As noted earlier in the context of calculating neutrino fluxes, because the steepness of the cosmic-ray spectrum, the effective average is $\sim  0.75$ times the maximum fraction. Also noting that the minimum fractional energy carried by the neutrino $\eta_{min}\approx 0$, $\eta_\mu\approx 0.75$, and is nearly a constant independent of the energy of the pion and the value of $\alpha$. The critical energy is $\varepsilon_{\eta\mu}=Hm_\mu/c\tau_\mu$.

The mean lifetime of the muon $\tau_\mu$ is $\sim 2.2\times 10^{-6} s$ so that even at $\sim1\textup{ } GeV$ its decay length is $\sim 6$ km, roughly equal to the scale height of the atmosphere. Thus in the calculation of the spectral intensities of the muons in the atmosphere at energies greater than about $10\textup{ } sec\textup{ }\theta$ GeV, we may safely neglect the decay of the muon. We may also neglect the energy losses due to ionization at $E > 30\textup{ } sec\textup{ }\theta$ GeV. Thus the muon intensity, $f_\mu(E,\textup{x}_{max},\theta)$, at the surface of the earth is given by the integral of the source function.
\begin{eqnarray}
\label{eq:45}
f_\mu(E,\textup{x}_{max},\theta ,\alpha )&=&\int_{0}^{\textup{x}_{max}} q_\mu(E,\textup{x},\theta ,\alpha ) d\textup{x}\nonumber\\
&=&A_\pi \lambda _\pi  \eta _{\mu}^{\gamma+1 }  \left \{ \frac{\eta _{\mu} \varepsilon _{\eta_\mu} (\alpha, \theta, E/\eta_\mu)}{\eta _{\mu} \varepsilon_{\eta_\mu} (\alpha, \theta, E/\eta_\mu)+E}\right \} 
\end{eqnarray}
In writing Eq.(~\ref{eq:45}) we have taken $\left \{ 1-\textup{exp}(\textup{x}_{max}/\lambda _\pi) \right \}\approx 1$.

\section{Comparison of the theoretical spectral intensities with cosmic ray observations}

\begin{figure}[ht]
\includegraphics[width=10cm]{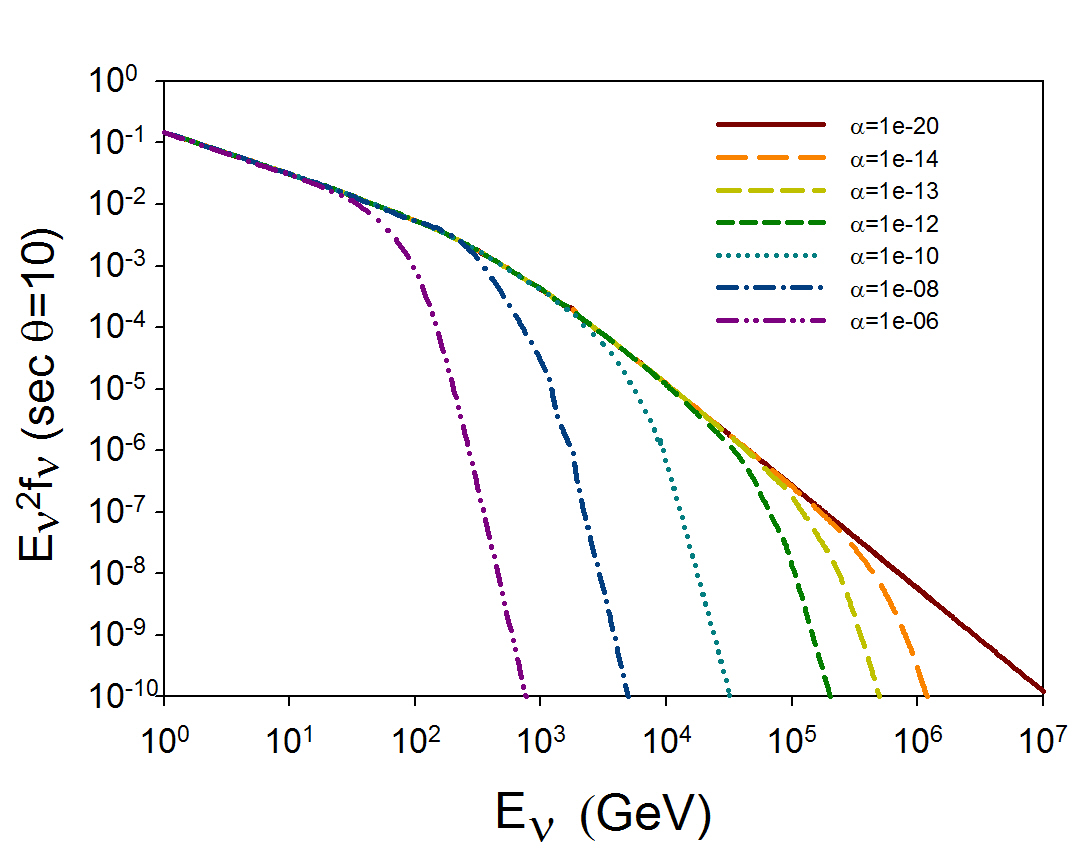}
\caption{The spectrum of neutrinos for various values of $\alpha$  and a fixed value of sec $\theta=10$ is displayed. Notice that the spectrum steepens sharply at progressively lower energies for increasing values of $\alpha$. For the extremely small value of $\alpha=10^{-20}$ there is no perceptible steepening even up to $10^7$ GeV.}
\label{fig:NeutVA}
\end{figure}
\begin{figure}[ht]
\includegraphics[width=10cm]{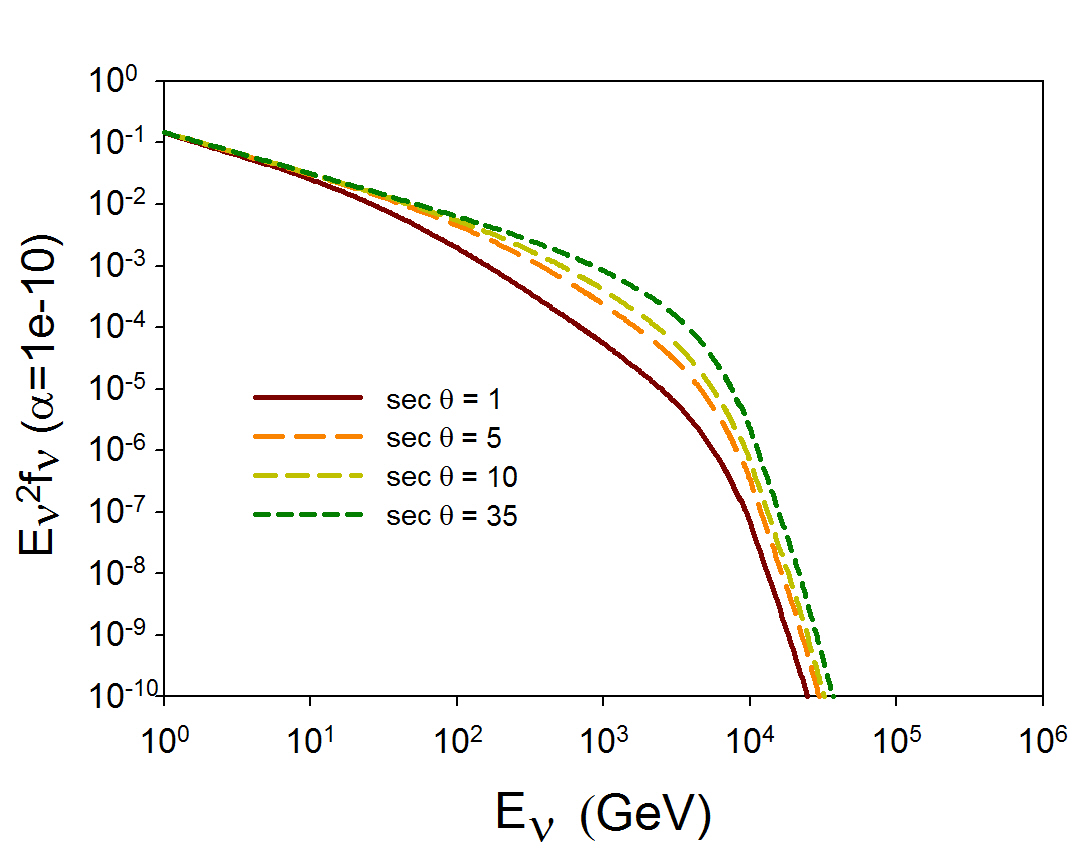}
\caption{The dependence of the neutrino spectra on sec $\theta$, for a fixed value of $\alpha=1\times10^{-10}$. The enhancement in the intensities at high energies with sec $\theta$ is seen, even with the presence of the LIV effects.}
\label{fig:NeutVS}
\end{figure}
\begin{figure}
\includegraphics[width=10cm]{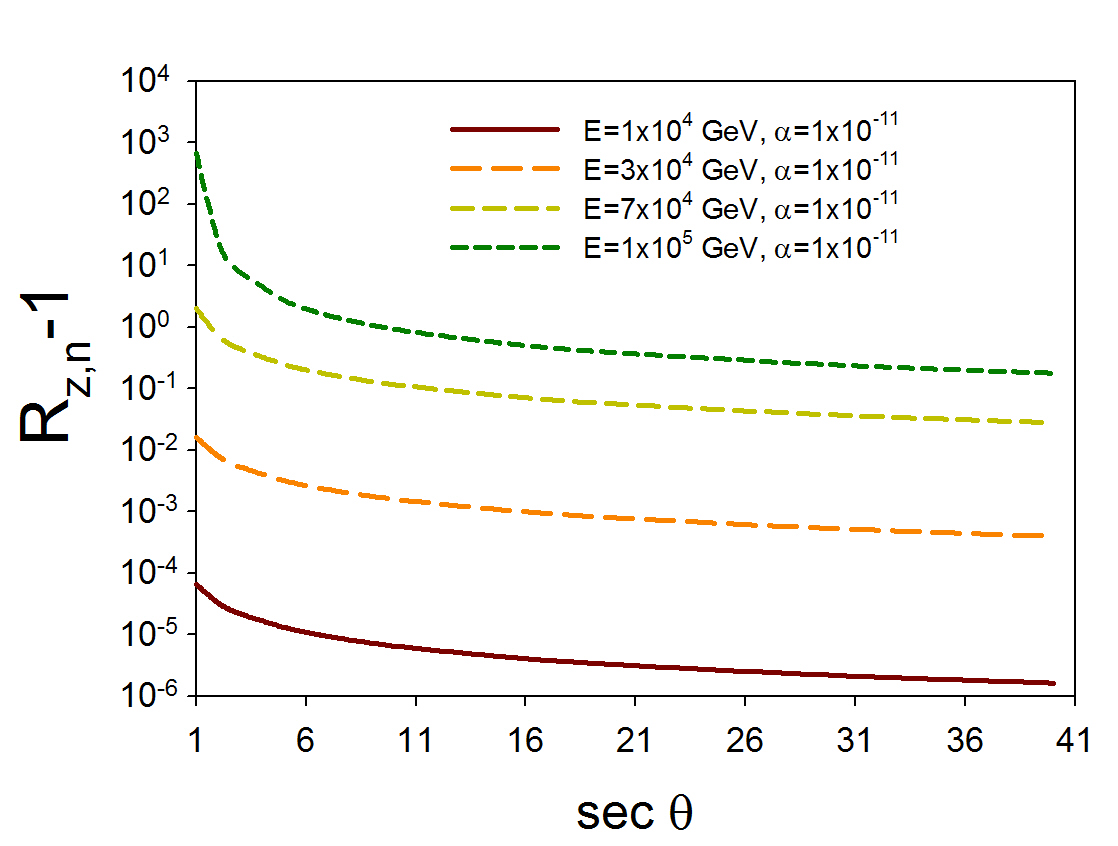}
\includegraphics[width=10cm]{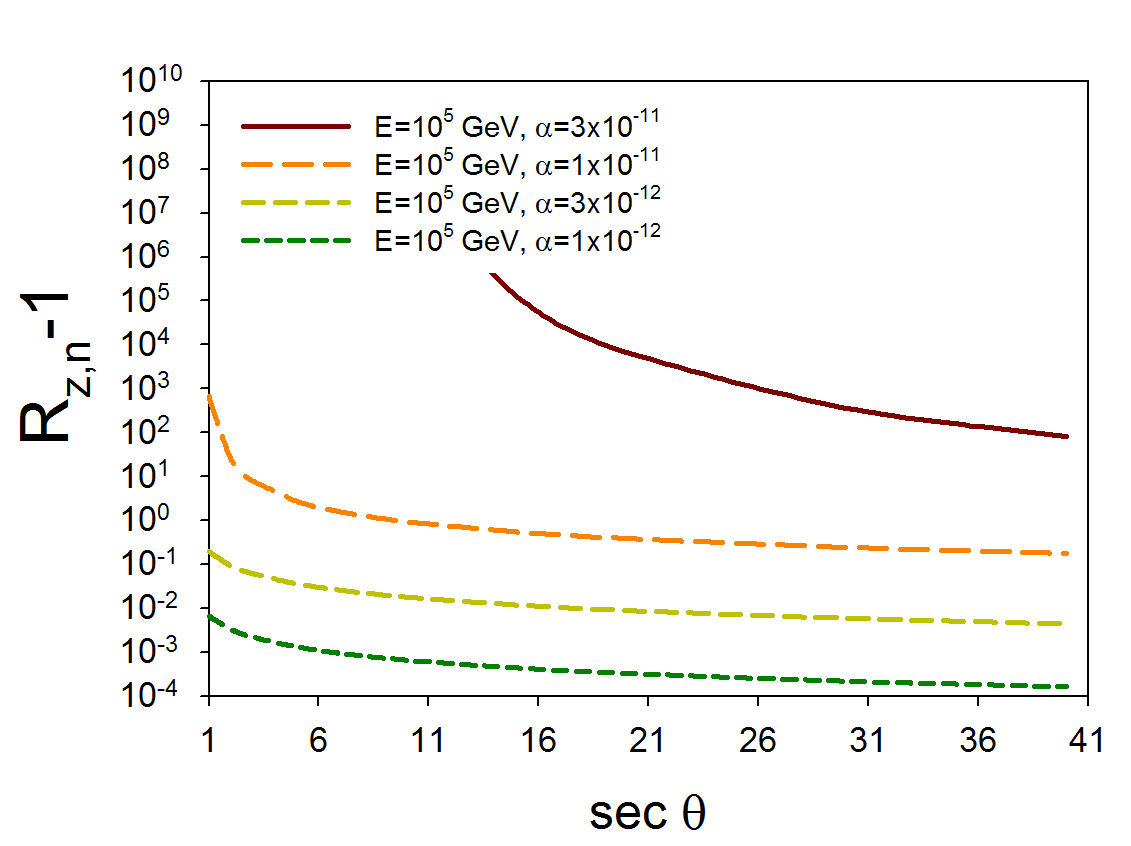}
\caption{The asymmetry in the spectral intensity in the forward/backward direction is displayed as a function of the arrival direction of the neutrino-induced muons in detectors placed $\sim 1$ km underground, for various values of the LIV parameter $\alpha$ (lower panel). Note that we have shown $R_{z,n}-1$ along the y-axis, in order to clearly bring out the dependence on the parameters, (see Eq. \ref{eq:43} in the text).}
\label{fig:RznVE}
\end{figure}
\begin{figure}[ht]
\includegraphics[width=10cm]{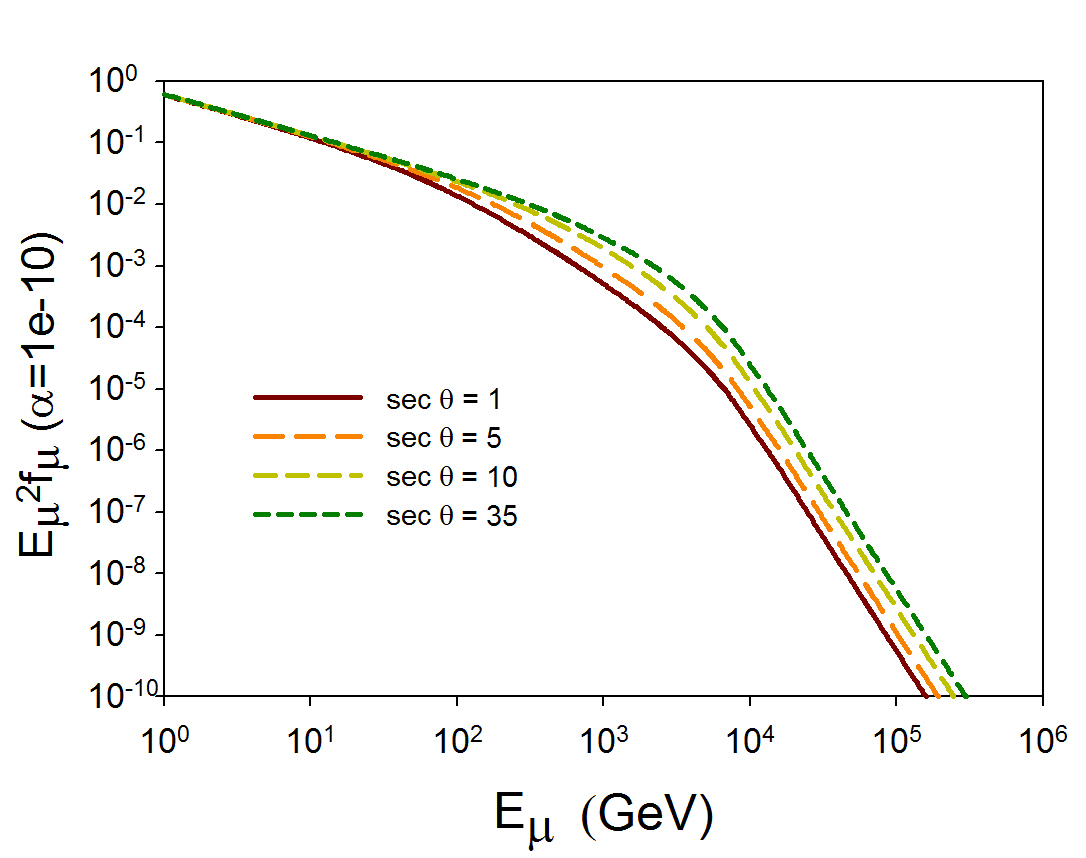}
\caption{This figure illustrates the muon spectra generated by the pions in the atmosphere with their decay times elongated by LIV effects; the sec $\theta$ enhancement of the intensities at high energies, well known in the cosmic-ray field is reproduced, even when LIV effects are present.}
\label{fig:MuonVS}
\end{figure}
\begin{figure}[h]
\includegraphics[width=10cm]{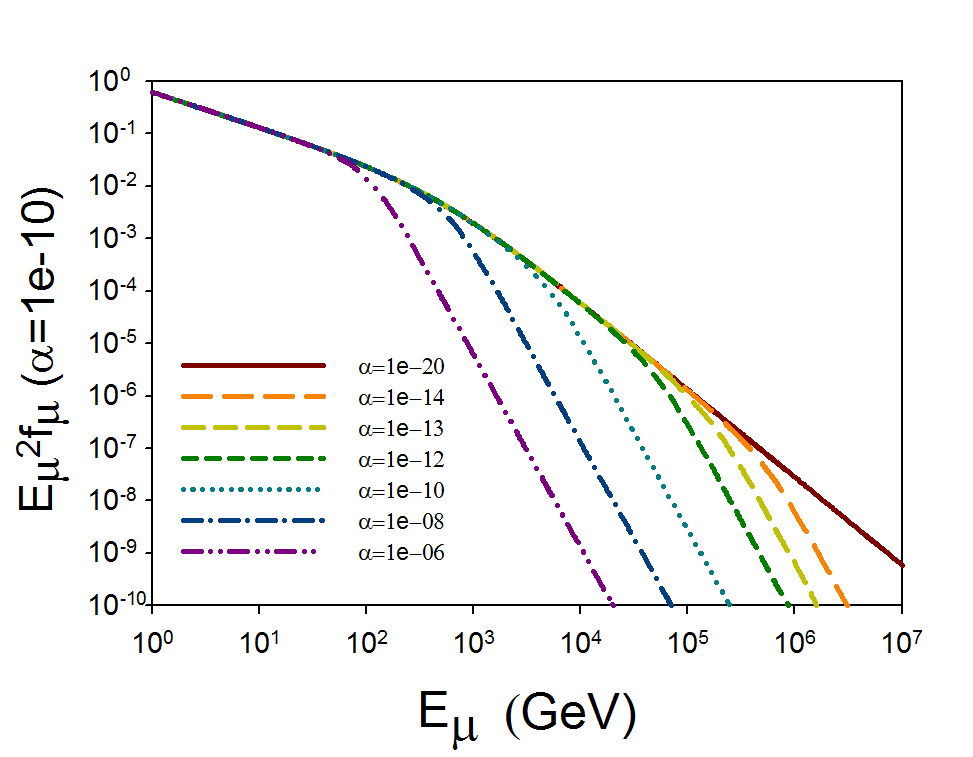}
\caption{The theoretically calculated muon spectra for various values of $\alpha$, with a fixed zenith angle, sec $\theta=10$, are displayed. The elongation of pion lifetime due to LIV effects make them preferentially interact in the atmosphere rather than decay. The progressive reduction of the high energy flux of muons with increasing $\alpha$ is seen clearly. }
\label{fig:MuonVA}
\end{figure}
\begin{figure}[h]
\includegraphics[width=12cm]{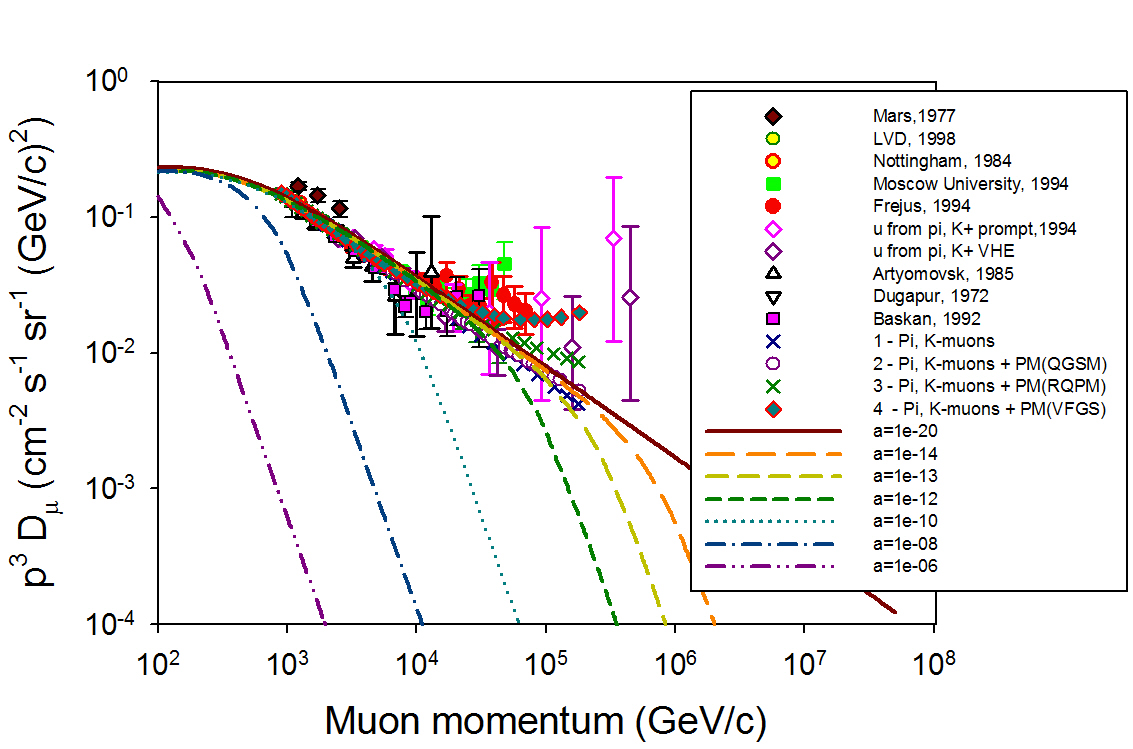}
\caption{Differential muon energy spectra from various experiments as reviewed by the IceCube collaboration \protect\cite{Novoseltsev2011} is compared with the theoretical expectation for different values of $\alpha$ and a fixed value of sec $\theta=1$. The theoretical spectra are normalized at $\sim1000$ GeV. Note that for $\alpha=10^{-13}$ the theoretical spectrum steepens significantly below the the observed intensities at $\sim10^5$ GeV. The theoretical spectra shown here are for a fixed sec $\theta=5$.}
\label{fig:MwDataVA}
\end{figure}

\noindent The calculations presented in the preceding section indicate that the spectral intensities of neutrinos and muons generated in the Earth's atmosphere through pion decay are sensitive to the posited level of LIV in the maximal attainable velocities of $\nu_\mu$. We illustrate the main effects of the superluminal motion of $\nu_\mu$ on the penetrating components of cosmic rays in a sequence of figures:  Figs. \ref{fig:NeutVA} 
- \ref{fig:MwDataVA}. The effect of increasing $\alpha$ on the spectrum of neutrinos is shown in Fig. \ref{fig:NeutVA}, where the value of sec $\theta$ is fixed at 10, and $\alpha$ is varied in the interval $10^{-20}$ to $10^{-6}$. Here we can see the neutrino spectra steepening from progressively lower energies with increasing $\alpha$. For the smallest value of $\alpha=10^{-20}$, there is no perceptible steepening even up to $\sim10^7$ GeV. In Fig. \ref{fig:NeutVS}, we show the dependence of the neutrino spectra on sec $\theta$, for a fixed value of $\alpha$. The well-known enhancement of the intensities at high energies with sec $\theta$, due to the increased fraction of pions which interact rather than decay, is reproduced even when the superluminal effects are included. The propagation of neutrinos through the Earth is exclusively determined by the Cohen-Glashow process, to the extent the neutrino interactions with the material of the Earth may be neglected or accounted for. The ratio of the neutrino spectral intensities $R_z(E,\theta,\alpha) $  at $\theta$ and $\theta+\pi$ calculated in Eq. (\ref{eq:43}) are displayed in Fig. \ref{fig:RznVE}. The $\theta$ dependence  of the ratio spectral intensities is displayed for $\alpha=10^{-11}$ for a set of neutrino energies $E=1\times 10^4$ to $1\times10^5$ GeV in the top panel and ratio for a fixed $E=10^5$ GeV for selected values  of $\alpha$ is shown as a function of energy in the bottom panel of Fig. \ref{fig:RznVE}.

Similarly we show in Fig. \ref{fig:MuonVS} the dependence of the muon spectra on sec $\theta$, for a fixed value of $\alpha$, and in Fig. \ref{fig:MuonVA} we show how the muon spectra become steep from progressively lower energies as we increase the value of $\alpha$, for a fixed value of sec $\theta$. We recall that our calculation neglects ionization and other losses of energy suffered by muons as they propagate through the Earth's atmosphere. The effects of such losses will be to flatten the spectra of muons at low energies.
We now proceed to compare these theoretical estimates with the available data and derive the bounds on the LIV parameter, $\alpha$. The first cosmic-ray observations of muon neutrinos date back to half a century or more, and the instruments have progressively increased in collecting power to achieve good sensitivities that we can observe cosmic ray neutrinos even up to $\sim$10$^6 $ GeV. Since the LIV effects in the Coleman-Glashow model increase with increasing energy, these observations probe very sensitively the effects of such violations.

\subsection{Comparison with cosmic ray neutrino intensities.}
\begin{figure} [h]
\includegraphics[width=11.6cm]{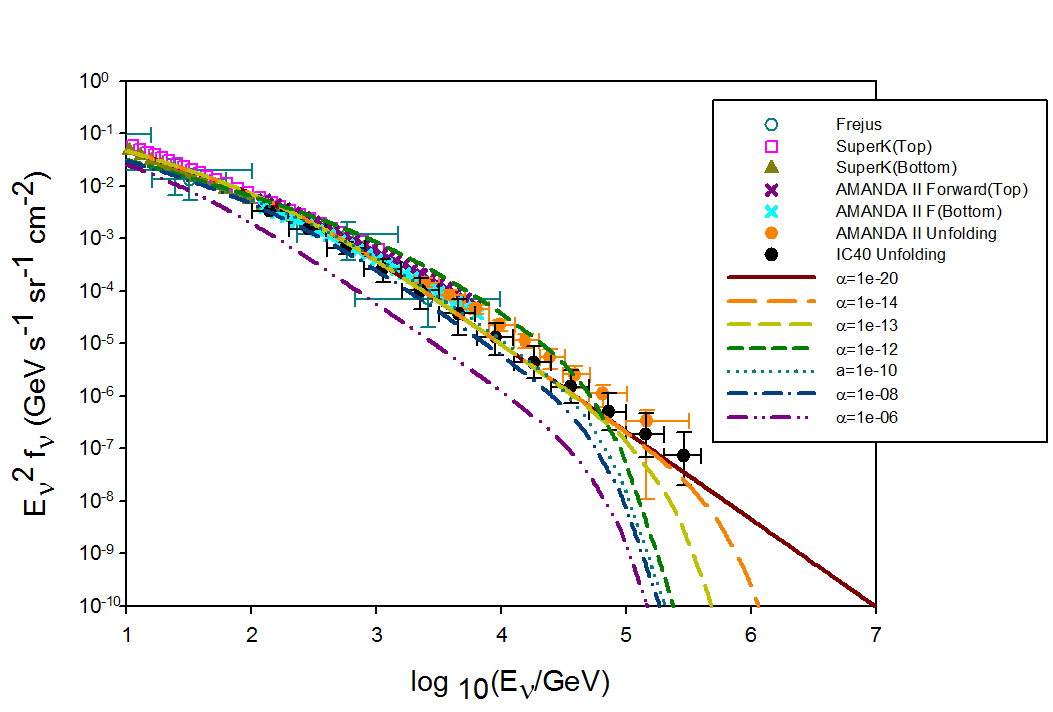}
\includegraphics[width=11.6cm]{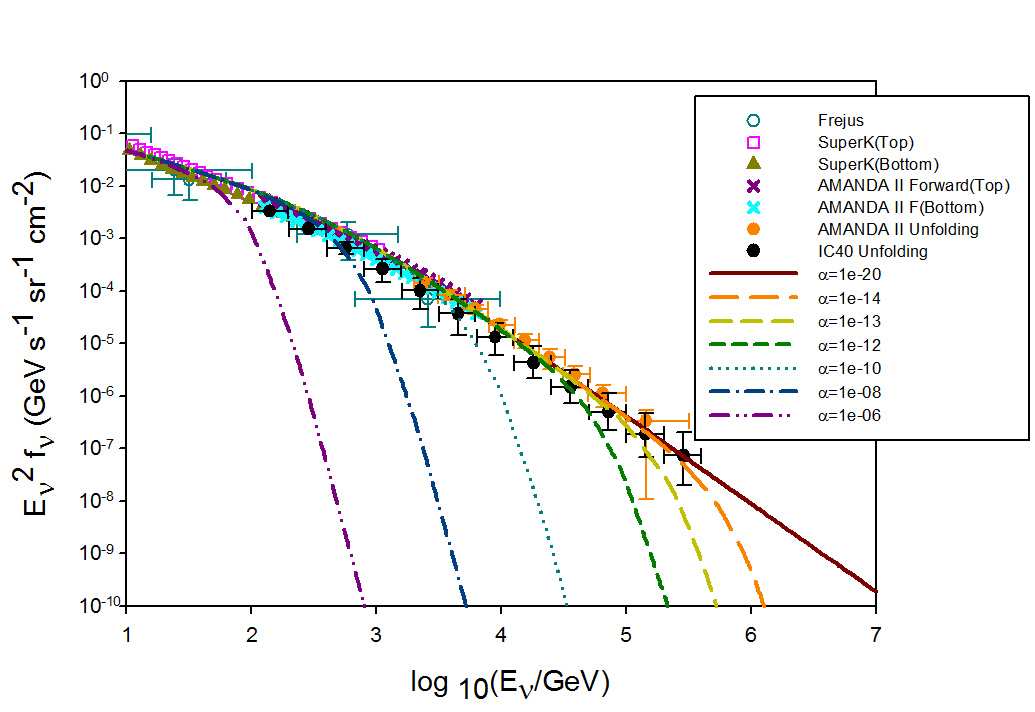}
\caption{Top panel: The theoretical spectral intensities of the muon neutrinos for various values of $\alpha$ are superimposed on the observed spectra for sec $\theta=5$, which is taken to represent the weighted average of the neutrino spectra at various zenith angles. Note that for very small $\alpha=10^{-20}(\approx0)$ the theoretical fluxes are in good agreement with the observations. Significant deviations appear for $\alpha \approx10^{-13}$ at high energies. Bottom panel: The value of sec $\theta$ has been changed to 10.}
\label{fig:NwDataVA}
\end{figure}
\noindent The measurement of cosmic ray neutrino fluxes started with the pioneering efforts of Reines \textit{et al.} \cite{Reines1965} and Achar \textit{et al.} \cite{Achar1965}. Progressively the size and sophistication of the detectors have improved so much that today we have the spectrum well measured by the IceCube collaboration up to $\sim3\times10^5$ GeV \cite{Abbasi2011}. The observed spectral intensities of neutrinos as reviewed by IceCube  \cite{Abbasi2011} is displayed in Fig. \ref{fig:NwDataVA}. In the same figure we superimpose the theoretical spectra calculated by us for various values of the LIV parameter $\alpha$, with sec $\theta$ = 5 representing the weighted average of the intensities over zenith angles of $90^{\circ}$ to $180^{\circ}$. In fact, the effective average value of sec $\theta$ increases with increasing neutrino energy as a consequence of the competition between interaction and decay of the pions. At the highest energies the mean value of sec $\theta$ is expected to be higher. For comparison, we show in the lower panel of Fig. \ref{fig:NwDataVA} the theoretically calculated neutrino intensities at sec $\theta=10$, along with the observational data. We normalize all the theoretical spectra to the observed fluxes at neutrino energies $\sim$500 GeV. 

We note that the theoretical spectra for $\alpha = 10^{-20} \approx 0$ fits the observations well. As $\alpha$ increases above $10^{-14}$, the theoretical curves start falling below the observation at the highest energies. For $\alpha=10^{-13}$, the theoretical curves fall a factor of $\sim300$ below the observed intensities at $\sim2\times10^5$ GeV, and by progressively smaller factors at lower energies. Thus it is safe to conclude that the value of the LIV parameter is less than $\sim10^{-13}$.
\subsection{Comparison with cosmic ray muon spectra.}
\noindent Novoseltsev \cite{Novoseltsev2011} provides an extensive compilation of the spectral intensities of cosmic ray muons at energies beyond 1 TeV, and we compare the theoretical spectrum given in Eq. (\ref{eq:45}) with data in Fig. \ref{fig:MwDataVA}. First, we note that for the very small values of $\alpha \approx 0$ the theoretical curve reproduces the observations very well, up to $\sim 2\times10^5$ GeV. The two points at $\sim4-5\times10^5$ GeV with large error bars lie above the theoretical predictions. For values of $\alpha>10^{-13}$, the theoretical curves peel off downwards from the data at progressively lower energies. We may thus conclude that these data also constrain the LIV parameter, $\alpha$ to be less than $\sim 10^{-13}$.

\section{Discussion}

\noindent In the preceding sections our aim was to provide a transparent description of the propagation of high energy cosmic rays in the atmosphere, resulting in simple analytical expressions for the spectral intensities of the neutrinos and muons. Despite the many simplifying assumptions made, our analytical expressions reproduce with adequate accuracy the observations and the well-known sec $\theta$ enhancement of the fluxes of the penetrating component at high energies. These analytical expressions incorporate the effect of novel LIV physics into the propagation of cosmic rays through the Earth's atmosphere and the propagation of neutrinos arising from pion decay through the Earth up to the detectors deployed deep underground.

By tracking how the spectral and angular dependence of the muons and neutrinos are thereby modified we have placed a strong limit on the LIV parameter: $\alpha< 10 ^{-13}$. Several remarks are now in order:
\newline
\noindent (i) \textsl{The electronic decay mode.}

\noindent The electronic $(e+\nu_e)$ decay mode of the pion is suppressed in the Lorentz Invariant standard model by a ``helicity" factor of $10^{-4}$. Furthermore unlike the muons, the energetic electrons generated by the cosmic ray $\nu_e$ in the Earth cannot penetrate very far from their production site. Thus our analysis and the bounds obtained with an exclusive focus on the muon sector do not depend on wether we include or ignore the electronic decay mode.
\newline
\noindent (ii) \textsl{Neutrino mixing.}

\noindent It has been pointed out \cite{Fargion2012} that the observed oscillations between neutrinos of different flavors are strongly suppressed if the LIV parameters $\alpha(\nu_\mu)$ and $\alpha(\nu_e)$ differ by more than $\Delta m_{1,2}^2/E(\nu)^2 \sim10^{-18}$ for $E(\nu)\sim$ O(10 MeV) and $\Delta m_{1,2}^2\sim10^{-4}$ eV$^2$. Assuming however an energy independent common - LIV ${\alpha}(\nu_i) \lesssim 10^{-10}$ seems quite consistent with all other data in particular the supernova 1987 upper bound of $\sim10^{-8}$ on $\alpha(\nu_e)$ \cite{Stodolsky1988, Longo1987}.
\newline
\noindent (iii) \textsl{Charm production.}

\noindent A $10^6$ GeV neutrino in cosmic rays would arise from the decay of pions of energy $\sim4\times10^6$ GeV, which in itself will be generated in the interactions of nucleons of $\sim4\times10^7$ GeV. Such an energy for the nucleons correspond to beam energies of $\sim5$ TeV in a collider. Accordingly, we may expect good data  for the production cross section for the very short lived charmed and other mesons which may decay readily giving high energy neutrinos. We may expect that such processes may start to dominate the neutrino fluxes at these energies \cite{Volkova1987, Volkova2009, Amsler2008}, making the spectra themselves not so very good probes of the $\alpha$ parameter. On the other hand these neutrinos will suffer energy losses through the Cohen-Glashow process and the asymmetry parameter $R_{z,n}(E)$ given in equation Eq. \ref{eq:43} will provide a useful signature of the superluminal neutrinos even in this energy region.
\noindent (iv) \textsl{LIV for charged leptons.}

\noindent Strict SU(2)$_L$ gauge invariance in the Standard Model suggests that any non-vanishing  $\alpha(\nu_i)$ parameter be associated with an equal LIV parameter for the corresponding lepton. Specifically we will then have: 
\begin{equation}
E(l_i) = [ m(l_i)^2 + (1+\alpha(\nu_i))^2 p_i^2]^{1/2} 
\label{eq:46}
\end{equation}
This would imply that asymptotically $E_i\sim(1+\alpha_i)p_i$ so that the muon also becomes equally superluminal at high energy. This, in turn, makes prolongation of the lifetime of the pions even much more dramatic - since at energies greater than $m_\pi/{\alpha^{1/2}}$ the pion becomes stable and neutrinos or muons of this energy should simply not be produced at all by high energy cosmic rays. Note however that if we extend this to a universality of all the $\alpha(i)$ including those of the electrons, then the Cohen-Glashow process is kinematically forbidden. An order of magnitude estimate indicates that $10^{-10}$ LIV alpha is (marginally) consistent with the precise measurements and calculations of the g-2 of the electron and certainly is allowed for the muon and the tau leptons.

When the Cohen-Glashow process is suppressed, then the leading process would be 
\begin{equation}
\label{eq:47}
\nu_\mu\rightarrow \nu' + \gamma 
\end{equation}
which will have a lower rate due to W in the loop. However, the severe limits on this process placed by Cowsik, Rajalakshmi and Sreekantan \cite{Cowsik1999b} will apply.

\noindent (v) \textsl{Connection to GRB's and Supernovae}

\noindent A direct $10^{-8}$ upper bound on the superluminality parameter $\alpha(\nu_e)$ is derived from the observed difference in arrival time of about a few hours between the neutrino pulses from SN1987a and the optical signature which travel together from the LMC to earth for $\sim3\times10^{12}$ seconds \cite{Stodolsky1988}. Gamma ray bursts (GRB's), detected by satellites at a rate of about one per day seem to originate at cosmological distances of $\sim$500 mega-parsecs $\sim10^4$ times larger than the distance from the 1987a supernova. Several GRB models suggest that protons should be accelerated up to $\sim10^{-20}$ eV in the fireball and the interaction of these energetic protons with the ambient material and radiation should generate pions which eventually decay into neutrinos. This motivated the IceCube collaboration to search for coincidences between GRBs and upward moving muons - namely energetic muons pointing in the same direction \cite{Abbasi2012}. The good angular resolution ($\sim$1 degree) of both IceCube and the satellite detectors make for small $\sim0.03$ probabilities of one random coincidence even within a generous time window of an hour. While no such coincidence has been observed to-date, future observations of any coincidences would lead to a $10^4$ times stronger bound $\alpha(\nu_{\mu}) < 5\times10^{-12}$, on the superluminality of muon neutrinos in the TeV range to be compared with that found for 10MeV electron neutrino by using SN1987a data. The bounds set by observing  neutrinos in coincidence with gamma ray bursts would improve by a further $\sim$100 for a coincidence observed within $\sim$1 minute. Amusingly, even using cosmological baseline in putative direct timing experiments yields bounds similar to those obtained by our analysis of high energy atmospheric muons and neutrinos! 

\noindent (vi) \textsl{G.Z.K and Anita UHE neutrinos.}

\noindent Since our bounds improve with the observation of higher energy neutrinos, the searches for ultra high energy  neutrinos with energies $\gtrsim10^9$ GeV are of special interest. The searches for such neutrinos were launched a while ago. The ANITA experiment suspended over Antarctica from a high altitude balloon detects the Cherenkov radio emission from neutrinos that skim the south pole ice \cite{Hoover2010}. So far only one candidate event has been found \cite{Gorham2010} and in view of a similar expected background, we cannot establish the existence of such ultra high energy neutrinos. Should however future flights provide enough statistics and neutrino initiated events of such high energies be definitely detected - the bounds will dramatically improve. To see this let us assume that the detected neutrinos are indeed of the GZK type namely coming from decays of pions produced via interactions of ultra high energy cosmic rays's with the background photons. The neutrinos would then have to travel some large distance $L\sim100$ Mega-parsecs $\sim10^{26}$ cm to arrive here. Demanding that the mean free path for energy loss through the Cohen-Glashow process of these neutrinos exceed this distance implies $E^5\alpha^3 < 10^{-26}$ so that $\alpha<10^{-23}$!!

\section{Conclusions}
\noindent In this paper we have developed a transparent analytical model for the propagation of cosmic rays in the Earth's atmosphere that explicitly includes the effects of a superluminal motion of the muon neutrino on the decay probability and kinematics of pion decay and energy losses suffered by the neutrino through the Cohen-Glashow process as it propagates from the production site in the atmosphere onwards to the proximity to the detectors placed deep underground. The available observational data to date on the cosmic ray generated neutrinos and muons place a bound of $\alpha\lesssim10^{-13}$. We have pointed out how detectors like IceCube may search exclusively for signatures of the Cohen-Glashow process by observing the forward-backward ratio of high energy neutrinos arriving at the same zenith angles but from upper and lower hemispheres. Keeping in mind that the data published by the IceCube collaboration is limited to those acquired in 2009 and earlier years, their full data up to the present date should be able to extend the spectra to $\gtrsim10^6$ GeV and the bounds on $\alpha$ to $\sim10^{-14}$. The observations of GZK neutrinos with $E \sim 10^9$ GeV will push $\alpha$ to well below $\sim10^{-23}$!!


\end{document}